\documentclass[twocolumn,showpacs,amsmath,amssymb,superscriptaddress,nofootinbib]{revtex4}

\usepackage{graphicx}
\usepackage{epsfig}
\usepackage{dcolumn}
\usepackage{bm}
\usepackage{color}

\newcommand{\beq}{\begin{equation}}
\newcommand{\eeq}{\end{equation}}
\newcommand{\bea}{\begin{eqnarray}}
\newcommand{\eea}{\end{eqnarray}}

\begin{document}
\title{Boost-invariant mean field approximation and the nuclear Landau-Zener effect}
\author{Lu Guo}
\affiliation {Institut f{\"u}r Theoretische Physik, J. W. Goethe-Universit{\"a}t, 
D-60438 Frankfurt, Germany}
\author{J. A. Maruhn}
\affiliation {Institut f{\"u}r Theoretische Physik, J. W. Goethe-Universit{\"a}t, 
D-60438 Frankfurt, Germany}
\author{P.-G. Reinhard}
\affiliation {Institut f{\"u}r Theoretische Physik $\mathit{II}$, Universit{\"a}t
Erlangen-N{\"u}rnberg, Staudtstrasse 7, D-91058 Erlangen, Germany}
\date{\today}

\begin{abstract}
We investigate the relation between time-dependent Hartree-Fock (TDHF)
states and the adiabatic eigenstates by constructing a boost-invariant
single-particle Hamiltonian. The method is numerically realized within
a full three-dimensional TDHF which includes all the terms of the
Skyrme energy functional and without any symmetry restrictions.  The
study of free translational motion of a nucleus demonstrates the
validity of the concept of boost-invariant and adiabatic TDHF
states. The interpretation is further corroborated by the test case of
fusion of $^{16}{\textrm O}$+$^{16}{\textrm O}$.  As a first
application, we present a study of the nuclear Landau-Zener effect on
a collision of $^{4}{\textrm {He}}$+$^{16}{\textrm O}$.
\end{abstract}

\maketitle

\section{\label{level1} Introduction}

The time-dependent Hartree-Fock (TDHF) approximation, originally proposed by
Dirac~\cite{Dirac-tdhf}, has found widespread applications in nuclear
dynamics since more than thirty years~\cite{Bonche76}.  It provides
the microscopic foundation for describing various dynamical scenarios
in the regime of large amplitude collective motion. Soon after its
introduction to nuclear physics, the TDHF approximation was
extensively applied to studies of fusion excitation functions,
fission, deep-inelastic scattering, collective excitations and nuclear
molecular resonances. These studies shed light on several mechanisms
of heavy-ion collisions and resonance dynamics. Reviews on these
earlier TDHF applications can be found, e.~g., in
Ref.~\cite{Negele82,Davies85}.

At that time, however, limited computer capacity restricted most
calculations to axial symmetry and omission of spin-orbit
coupling. These limitations turned out to be a hindrance for the
development. For example, earlier TDHF calculations underestimated the
energy dissipation from the collective kinetic energy into internal
excitations so that the energy window of fusion reactions was too
small in comparison with experiments.  Later work demonstrated that
the spin-orbit coupling \cite{Reinhard88,Umar89} and fully
three-dimensional geometry~\cite{Davies78,Flocard78,Bonche78} enhance
dissipation. None of these calculations, however, was able to include
all constituents simultaneously.
With the steady upgrade of computational power, three-dimensional TDHF
calculations employing the full Skyrme force became possible and
renewed the interest in nuclear TDHF as seen from recent publications
on resonance dynamics~\cite{Simenel03,Reinhard07,Naka05,Umar05a,Maruhn05} and
heavy-ion collisions~\cite{Maruhn06,Umar06a,Umar06b,Simenel01}. One
expects that the new generation TDHF calculations may yield more
realistic features for heavy-ion collisions at low and medium energies
and for resonance dynamics.  This revives old questions that have been left
unanswered for a while, one of which is the subject of this paper:
the analysis of heavy-ion collisions as computed by
TDHF in terms of adiabatic states and level-crossing dynamics.

Nucleus-nucleus collisions present different behavior depending on the
delicate balance of reaction time and rearrangement time of the mean
field. For example, the experimentally observed resonance-like peaks
in the inelastic cross sections are explained in terms of the
well-known Landau-Zener excitation due to a breakdown of the
adiabaticity condition near an avoided level crossing.  The
Landau-Zener mechanism was first introduced into nuclear physics in
Ref.~\cite{Hill53}.  The Landau-Zener effect and its applications to
heavy-ion collisions have been discussed~\cite{Park80,Park82,Cas85a}
in terms of the asymmetric two-center shell model
(TCSM)~\cite{Maruhn72}, employing the assumption that the nucleons can
be described with adiabatic ``molecular'' states, i.~e. the
instantaneous eigenstates of the deformed mean field, during the heavy
ion reaction.  A review on Landau-Zener dynamics and experimental data
in nuclear molecules may be found in Refs.~\cite{Cindro86,Thiel90}.

TDHF calculations, however, have never been analyzed in terms of the
Landau-Zener effect because their single-particle states and energies
have no simple physical interpretation and a construction of
corresponding adiabatic states was not readily available. There were
early attempts to define a related adiabatic basis by means of
density-constrained Hartree-Fock~\cite{Cus85a,dencon}. These turned
out to be very promising, lacking, however, at that time the exact
treatment of the flow contributions. Improved computing power now
allows revisiting the case without technical restrictions. We thus
will present here a self-consistent scheme to define and compute two
useful analyzing instruments for a given TDHF state: instantaneous
single-particle energies and instantaneous adiabatic states. The
straightforward expectation values of the single-particle Hamiltonian
turn out to be blurred by trivial flow contributions.
To take into account the effect of motion on the single-particle wave
functions, we define a single-particle Hamiltonian which is invariant
under Galilei transformation, in particular under a boost. We call
that the boost-invariant Hamiltonian.  This dramatically reduces the
energy variances of the actual TDHF states, providing better-defined
single-particle energies.  Moreover, the (instantaneous) eigen-states
and energies of the boost-invariant Hamiltonian provide a well
defined adiabatic basis.  For example, the single-particle states in
free translational motion are exact eigenstates of this
boost-invariant Hamiltonian and its expectation values remain the
static single-particle energies. While this property is not exact
anymore in a situation of two colliding nuclei, we shall demonstrate
that this method nevertheless allows to define meaningful
single-particle energies and variances thereof.  Moreover, one can
establish an approximate relation of the time-dependent solutions to
the adiabatic deformation-dependent spectrum.  As a first application we
study the Landau-Zener effect in heavy-ion collisions.

The paper is organized as follows: Section \ref{level2} briefly
recalls the Skyrme energy functional and TDHF. In section \ref{level3}
we construct a boost-invariant Hamiltonian and test its validity for
free translational motion of a nucleus.  Section \ref{level4} applies
the newly developed scheme to the analysis of level crossings in
heavy-ion collisions. Section \ref{level5} is devoted to the summary and
conclusion.

\section{\label{level2} Comments on TDHF with Skyrme forces}

Most nuclear TDHF calculations by far are based on the Skyrme energy
functional; for a recent extensive review see~\cite{Bender03}.  It is
also used in the present applications. The starting point is the
Skyrme energy-density functional
${\cal{E}}_{Sk}={\cal{E}}(\rho,\tau,\vec{\sigma},\vec{j},\vec{J})$,
which is expressed in terms of a few local densities and currents:
density $\rho$, kinetic density $\tau$, spin density $\vec{\sigma}$,
current $\vec{j}$, and spin-orbit density $\vec{J}$. It includes free
kinetic energy, Skyrme interaction, Coulomb energy, and the
center-of-mass correction. The pairing energy is ignored here as we
will deal with collisions of closed-shell nuclei.
There are various parametrizations of the Skyrme
force~\cite{Bender03}. Since the present study is concerned with
fundamental effects which should not depend on the detailed force
used, we thus chose just one parametrization out of many, namely the
force SLy6~\cite{Chabanat98} which is widely used and provides a
reliable description of nuclear structure and dynamics.

Using the principle of least action and varying with respect to the
single-particle state $\varphi_\alpha^*$, we obtain the TDHF equations
(in the following units with $\hbar=1$ are used)
\begin{equation}
  {\rm i}\partial_t\varphi_\alpha 
  = 
  \hat{h}(\rho,\tau,\vec{\sigma},\vec{j},\vec{J})\varphi_\alpha
  \quad,
\label{tdhf}
\end{equation}
where $\hat{h}$ is the time-dependent mean-field Hamiltonian depending
on the occupied single-particle wave functions through densities and
currents. Given the initial conditions,
$\{\varphi_\alpha(\vec{r},t=0)\}$, the TDHF equations~(\ref{tdhf})
determine the wave functions for all later times.
In the stationary limit, we obtain the static mean-field equation
\begin{equation}
  \hat h\varphi_\alpha 
  = 
  \varepsilon_\alpha\varphi_\alpha   
\quad,
\end{equation}
where the single-particle energies $\varepsilon_\alpha$ appear
naturally as eigenvalues of the mean-field Hamiltonian $\hat h$.
The question is how to generalize the definition of the single-particle
energy to TDHF. The naive definition is to use the expectation value
of the instantaneous mean field $\hat{h}(t)$. This, however, raises
difficulties as we will see. Possible improvements will be developed
in the sequel.

A few words on the numerical solution are in order.  The set of non-linear TDHF
equations is solved on a three-dimensional Cartesian coordinate-space grid
employing a Fourier representation for the derivatives. All contributions of the
full Skyrme force were included and no symmetry restrictions imposed.  The
coordinate-space grid consists of $24\times24\times24$ points with a grid
spacing of 1~fm.  For the dynamical time stepping, we use a Taylor series
expansion of the unitary mean-field propagator up to sixth order
\cite{Flocard78} and a time step of 0.2~fm/$c$.  These numerical parameters
provide good conservation of particle number and total energy during the dynamic
evolution.  The static HF equations were solved with the damped gradient
iteration method~\cite{Rei82a,Blum92}.

\section{\label{level3} Single-particle energies in a moving frame}

\subsection{Adiabatic expansion as a propaedeutic example}

Adiabatic single-particle states are eigenstates of a single-particle
Hamiltonian for a given set of deformation parameters.  For a first
introduction, we will discuss that concept in this section on the
grounds of a given, properly parametrized, single-particle
Hamiltonian, such as, e.~g., that of the TCSM \cite{Maruhn72}. Such a
Hamiltonian usually depends on a few collective deformation parameters
which characterize the wanted reaction path. For simplicity let us
just deal with internuclear distance $R$ as the sole such parameter
and skip trivial complications such as spin and isospin. The proper
choice of an $R$-dependent single-particle potential $V(\vec r;R)$ and
the subsequent solution of the eigenvalue problem yield a set of
adiabatic single-particle states $\phi_k(\vec r;R)$ and corresponding
eigenenergies $\epsilon_k(R)$.
Using the adiabatic single-particle states the time-dependent (but
still independent-particle) solution is expanded as
\begin{equation}
  \psi_j(\vec r,t)
  =
  \sum_k c_{jk}(t)\phi_k(\vec r;R(t))
  e^{-{\rm i}\int_{t'}^tdt'\,\epsilon_k[R(t')]}
  \quad.
\label{adia}
\end{equation}
Such an expansion underlies, e.~g., the cranking
model~\cite{cranking1,cranking2}. This expansion has the problem that
it relies on a stationary basis in which the current vanishes for all
states. Any flow has to be described through the complex expansion
coefficients $c_{jk}$. This limits the ansatz (\ref{adia}) to
extremely slow motion.

An instructive example is uniform center-of-mass translation of an
unexcited nucleus with velocity $\vec{v}=\vec{p}/m$.  It is
described by coherently moving single particle states $\psi_k(\vec
r,t)= \phi_k(\vec{r}-\vec{v}t) \exp\left({\rm i}(\vec
k\cdot\vec{r}-\epsilon_kt)\right)$.  Clearly, the trivial plane-wave
factor which produces the correct flow is missing from the basis
states in (\ref{adia}). It has to be reconstructed laboriously by the
expansion coefficients.  A much more efficient description could be
obtained by properly extending the scheme to a dynamic basis
$\phi_k(\vec r;R,\dot{R})$ which already accounts for collective
flow. This step was found to be crucial for the derivation of
microscopic theories for collective motion in the framework of
adiabatic TDHF \cite{Goe78a}. We will now use the extension for the
definition of adiabatic reference states in TDHF. The example of
center-of-mass motion will be used as guidance.

\subsection{Flow-induced variances}

We will now discuss the case of self-consistent mean fields.  To simplify
the formal considerations, we restrict the discussion to one
spatial dimension and think in terms of the simplest energy functional
${\cal E}={\cal E}(\rho)$ depending only on the local density and
producing a purely local mean field, i.~e.,
\begin{equation}
  \hat{h}[\rho]
  =
  \frac{\hat{p}^2}{2m}
  +
  U(x,t)
  \quad,\quad
  U(x,t)
  =
  \frac{\delta{\cal E}}{\delta\rho(x,t)}
  \quad,  
\end{equation}
where $U$ is obtained from ${\cal E}$ by functional derivation. 
The ground-state wave functions $\{\varphi_{0,\alpha}\}$ fulfill the equations
\begin{equation}
  (\hat{h}_0-\varepsilon_\alpha)\varphi_{0,\alpha}(x)
  = 
  0
\label{stamf}
\end{equation}
with $\varepsilon_\alpha$ being the static single-particle energies.

As a test case, consider center-of-mass motion of the HF ground
state.  The motion is initialized by a boost with total momentum
$P=MV$ where $V$ is the velocity of the center of mass and $M=Nm$ the
total mass. The same boost is applied to all single-particle
wave functions
\begin{subequations}
\begin{eqnarray}
  \varphi_{0,\alpha}
  &\longrightarrow&
  \varphi_\alpha(x,t)
  =
  e^{{\rm i} P\hat{x}/N}\varphi_{0,\alpha}(x-Vt)
  e^{-{\rm i}\tilde\varepsilon_\alpha t},
\label{eq:boost1}
\\
  &&
  \tilde\varepsilon_\alpha 
  =
  \varepsilon_\alpha+\frac{P^2}{2M}
  \quad.
\label{eq:boost}
\end{eqnarray}
\end{subequations}
The local density is then propagated with velocity $V$ as
$\rho(x,t)=\rho_0(x-Vt)$ and this, in turn, carries through to the
mean field motion as
                 $U(x,t) = U_0(x-Vt)$.
The boosted wave functions together with the similarly moving mean
field are the solution of the TDHF equations~(\ref{tdhf}). The action of
the mean field on the boosted wave function can be expressed in terms
of the static solution as
\begin{eqnarray}
  \hat{h}\varphi_\alpha
  &=& 
  e^{{\rm i} P\hat{x}/N}e^{-{\rm i}\tilde\varepsilon_\alpha t}
  \big(\tilde\varepsilon_\alpha+\frac{P}{M}\hat{p}\big)\varphi_{0,\alpha}.
\label{eq:splitH}
\end{eqnarray}
The expectation value is simply 
$\langle\varphi_\alpha|\hat{h}|\varphi_\alpha\rangle
  =  \tilde\varepsilon_\alpha
$.
Both together allow evaluation of the variance of the single-particle
Hamiltonian explicitly as
\begin{equation}
  \langle\varphi_\alpha|\Delta\hat{h}^2|\varphi_\alpha\rangle
  = 
  \frac{P^2}{M^2}\langle\varphi_{0,\alpha}^{\mbox{}}|\hat{p}^2
       |\varphi_{0,\alpha}^{\mbox{}}\rangle .
\label{eq:variat}
\end{equation}
It is obvious that the moving wave function $\varphi_\alpha$ is not an
eigenstate of the instantaneous mean-field Hamiltonian $\hat{h}$. The
variance grows quadratically with the boost momentum $P$,
i.~e., proportionally to the center-of-mass energy. The expectation value
$\tilde\varepsilon_\alpha$ also becomes misleading. The kinetic
contribution makes the binding properties invisible.
This problem was already
noticed by Thouless and Valatin~\cite{Thouless62} 
while they were studying Galilean invariance of the TDHF equation.

\subsection{Construction of a boost-invariant mean field}

The above example of center-of-mass motion is instructive. The
variance of the mean-field Hamiltonian grows although we know that
the system remains intrinsically unaltered. All that happens is  a trivial
kinematical effect. Thus there should be ways to undo it equally
trivially. In the center-of-mass case, we could simply transform the
single-particle momentum into the intrinsic frame as 
$\hat{p}\longrightarrow\hat{p}-P/N$
and use that in the kinetic energy operator. That indeed provides a
reasonable boost-invariant Hamiltonian for that particular case. A
generalization can be obtained with the concept of the local momentum
distribution $\overline{p}(x)$ as given by the local current
$j(x)$. This suggests the definition of a locally boost-invariant
momentum 
\begin{equation}
  \hat{p}
  \longrightarrow
  \hat{p}_{\rm inv}
  =
  \hat{p}-\overline{p}(x)
  \quad,\quad
  \overline{p}(x)
  =
  \frac{j(x)}{\rho(x)}
  \quad,
\end{equation}
which can be extended to a boost-invariant kinetic-energy density
$
\tau_{\rm inv}
=
\sum_\alpha|\hat{p}_{\rm inv}\varphi_\alpha|^2
=
\tau-j^2/\rho
$.
It is interesting to note that this is practically the
Galilean-invariant combination $\tau\rho-j^2$ of kinetic contributions
in the interaction part of the Skyrme energy
functional~\cite{Bender03,Eng75a}. This gives confidence in the above
generalization.

The idea thus is to define an ``intrinsic'' energy-functional by
replacing the kinetic energy $\propto\int d^3r\tau$ by the
boost-invariant kinetic energy
\begin{equation}
 E_{\rm kin,inv} 
 = 
 \frac{1}{2m}\int dx \left(\tau-\frac{j^2}{\rho}\right)
 \quad.
\label{eq:invekin}
\end{equation}
The potential energy was already boost-invariant and thus the total
functional becomes so. This functional is to be used for the purpose
of analysis only, and it plays no role for the computation of the time
evolution as such.
Variation leads to the corresponding boost-invariant mean-field
Hamiltonian
\begin{equation}
  \hat{h}_{\rm inv} 
  = 
  \frac{\hat{p}^2}{2m}+U(x,t)
  - \frac{1}{2m}\big\{\frac{j(x)}{\rho(x)},\hat{p}\big\}
  + \frac{j^2(x)}{2m\rho^2(x)}\quad,
\label{eq:invH}
\end{equation}
where $\{...,...\}$ is the anti-commutator and $U(x,t)$ the usual
time-dependent mean-field potential.  The first two terms are exactly
the same as in the usual TDHF Hamiltonian and the last two stem from
the boost-invariant kinetic energy. The corrected Hamiltonian then
defines a boost-invariant single-particle energy
\begin{equation}
  \varepsilon^{\rm(inv)}_\alpha
  =
  \langle\varphi_\alpha|\hat{h}_{\rm inv}|\varphi_\alpha\rangle.
\label{eq:invspe}  
\end{equation}
Next we will show that the boost-invariant Hamiltonian has the boosted
TDHF wave functions as eigenstates in the case of free translation.

\subsection{Test case: free translational motion}

The construction of the boost-invariant Hamiltonian (\ref{eq:invH})
was guided by the example of free center-of-mass translation, so that the
natural test case is the global center-of-mass boost
(\ref{eq:boost1}).  In that case, the local momentum distribution
is given by
\begin{equation}
  \overline{p}(x)=\frac{j(x)}{\rho(x)}=\frac P{N}=\mbox{const.}
  \quad,
\end{equation}
while the boost-invariant Hamiltonian reduces to
\begin{equation}
  \hat{h}_{\rm inv}
  =
  \frac{\hat{p}^2}{2m}+U(x,t)-\frac{P}{M}\hat{p}+\frac{P^2}{2M}
  \quad,
\end{equation}
and finally the expectation value of the 
boost-invariant Hamiltonian and its variance become
\begin{subequations}
\begin{eqnarray}
\langle\varphi_\alpha|\hat{h}_{\rm inv}|\varphi_\alpha\rangle &=& \epsilon_\alpha,
\\
 \langle\varphi_\alpha|\Delta\hat{h}^2_{\rm inv}|\varphi_\alpha\rangle
  & = &
  0,
\label{eq:invHvariance}
\end{eqnarray}
\end{subequations}
where $\varepsilon_\alpha$ are the static single-particle energies.

The zero variance means that the boosted TDHF wave function
$\varphi_{\alpha}$ according to Eq. (\ref{eq:boost1}) is an eigenstate
of the boost-invariant Hamiltonian $\hat{h}_{\rm inv}$ and its eigenvalue remains
static solution $\varepsilon_\alpha$ as defined in Eq.~(\ref{stamf}).

These results for free translation suggest the boost-invariant
Hamiltonian $\hat{h}_{\rm inv}$ as an appropriate instrument for
analyzing the single-particle states of TDHF. The single-particle
energies computed as expectation values of the boost-invariant
Hamiltonian~(\ref{eq:invH}) can be considered as ``intrinsic''
single-particle energies representing the actual binding independent
from trivial kinematical contributions. 
It is to be noted that these equations 
are also applicable for a non-local mean-field
Hamiltonian like, e.g., for the Skyrme force.

The practical computation of the boost-invariant Hamiltonian is a bit
demanding due to the density in the denominator. Nevertheless, for
free translation of the nucleus $^{16}{\textrm O}$ we achieve
variances of about 0.02--0.05 MeV and expectation values stay within
$10^{-4}$$\sim$$10^{-5}$ MeV of the static ones with the
full three-dimensional TDHF which includes all the terms of the
Skyrme energy functional and without any symmetry restrictions. 

\subsection{Adiabatic states}

True intrinsic excitations in more general dynamical situations add
some energy variance to the TDHF states. The eigenstates of the
instantaneous boost-invariant Hamiltonian,
\begin{equation}
  \hat{h}_{\rm  inv}\phi_i
  =
  \varepsilon^{\rm(adia)}_i\phi_i
  \quad,
\end{equation}
then become something different. They correspond to some extent to the
adiabatic states and we will therefore use the expressions ``adiabatic
states'' and ``adiabatic energies'' in the following.  A given TDHF
state $\varphi_\alpha$ imbued with some intrinsic excitation is
distributed over the adiabatic basis $\left\{\phi_i\right\}$. This can
be quantified through the adiabatic occupation probability
\begin{equation}
  P_{i}^{\rm(occ)}
  = 
  \sum_{\alpha\in{\rm occ}}\left|\langle\varphi_\alpha|\phi_i\rangle\right|^2
\label{eq:adiabocc}
\end{equation}
where the sum runs over the occupied TDHF states $\varphi_\alpha$.
That quantity is the probability to find the state $\phi_i$ occupied
when expanding the actual TDHF Slater state into the adiabatic
basis. Complementarily we have the hole probability
$P_{i}^{\rm(unocc)}=1-P_{i}^{\rm(occ)}$.  The occupation probability
quantifies in its way the amount of intrinsic excitation carried in
the TDHF states. The value of one means no excitation at all 
for states of the adiabatic basis below the Fermi level, and lowering
below one is closely related to excitation.

\section{\label{level4} Analysis of heavy-ion collisions}
\subsection{Level schemes}

\begin{figure}
\epsfxsize=8.6cm 
\centerline{\epsffile{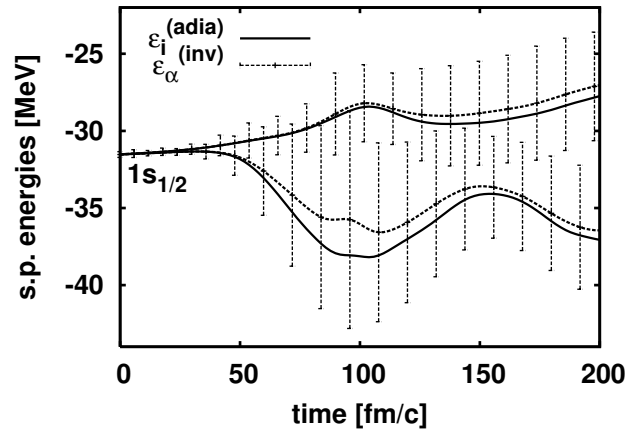}}
\caption{\label{vari} 
Time evolution of boost-invariant and adiabatic observables for a
central $^{16}$O+$^{16}$O collision at 25 MeV. The dashed lines with error bars
show the boost invariant single-particle energies $\varepsilon^{\rm(inv)}_\alpha$  
and their variances $\Delta\varepsilon^{\rm(inv)}_\alpha$ for the lowest ($1s_{1/2}$) proton 
states, and solid lines the adiabatic energies $\varepsilon^{\rm(adia)}_i$ 
for comparison. }
\end{figure}

\begin{figure*}
\epsfxsize=16.6cm 
\centerline{\epsffile{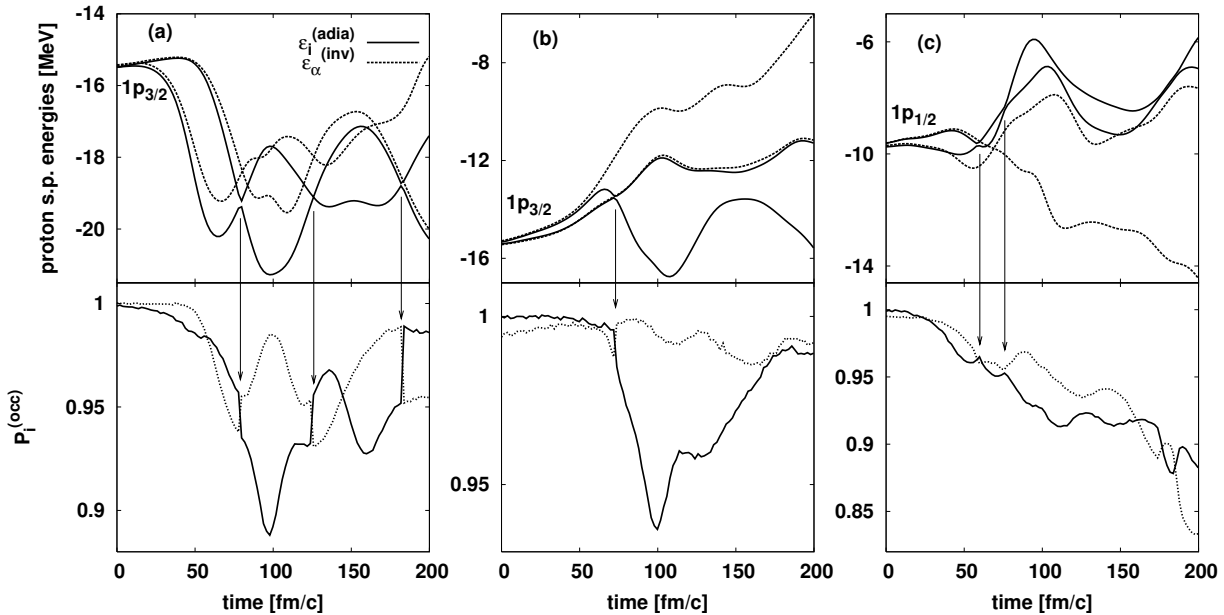}}
\caption{\label{spectrum} 
Time evolution of higher lying proton states for a central collision
of $^{16}\textrm {O}$+$^{16}{\textrm O}$ with a center of mass energy
of 25 MeV.  The upper panels show the boost invariant single-particle
energies $\varepsilon^{\rm(inv)}_\alpha$ (dashed) and the adiabatic
energies $\varepsilon^{\rm(adia)}_i$ (full lines). The lower panels show the corresponding
adiabatic occupation probabilities. The arrows pointing from the upper
panels to the lower ones indicate the level crossings where occupation
changes rapidly. The labels of the single-particle states at t=0 are 
indicated near the left axis.}
\end{figure*}

For the investigation of heavy-ion collisions, the wave functions of
the two fragments are placed symmetrically on the grid 5~fm off the
center of box and then boosted to the desired relative center-of-mass
energies.  This is the initial state for TDHF dynamical propagation.
Besides the calculation of the usual TDHF states
$\left\{\varphi_\alpha\right\}$ of the colliding system,
diagonalization of the boost-invariant Hamiltonian gives the adiabatic
single-particle energies and wave functions
$\left\{\phi_i\right\}$. It should be noted that in a collision
situation the boost-invariant Hamiltonian cannot be expected to
eliminate the kinetic effects completely, because it depends on the
{\em local} conditions. It is hoped, however, that for the initial
stages of the interaction the behavior of the levels can still be
extracted. 
For the symmetric system, parity projection has been done for the 
boost-invariant wavefunctions.

To illustrate the performance of boost-invariant and adiabatic states,
Fig.~\ref{vari} shows a typical result for the splitting of an
initially highly degenerate state during a heavy-ion reaction. The test
case is a central $^{16}\textrm {O}$+$^{16}{\textrm O}$ collision with
center-of-mass energy of 25 MeV. Shown are the two lowest levels (the
proton $1s_{1/2}$ levels in the right and left collision
partner).
The first few fm/c show the initial phase with nearly free c.m. motion
of the two nuclei towards each other. Boost-invariant and adiabatic
states remain identical and the energy variance is zero within the
limit of numerical precision.
The nuclei start to interact around 30 fm/c. At this time the
initially degenerate 1s$_{1/2}$ states split into two levels, as
expected.  The way the levels split into the various sub-states is
very similar to what is expected from a two-center approach like
in~\cite{Maruhn72}.
At the same time the energy variance of the boost-invariant states
grows because the collision mixes forward with backward flow such that
we have an increasing spread of flow around its decreasing average.
It is to be noted that the variances of the boost-invariant Hamiltonian are
much smaller than those of the usual TDHF Hamiltonian. The latter comes up
to several tens of MeV clearly showing that the single-particle
energies from the TDHF states have no meaning at all.
Furthermore, the boost-invariant energies become slightly larger than
the adiabatic ones, which expresses the amount of true intrinsic
excitation piling up in the boost-invariant states. It is very
satisfying to see that the energy difference is proportional to the
variance of the boost-invariant states. The signals confirm each other
as measure for intrinsic excitation and they give credibility to both
forms of energy expectation values.
Altogether, figure~\ref{vari} clearly demonstrates the usefulness of
the concept and the relation between the boost-invariant states
containing local flow and the quasi-stationary adiabatic states.

Figure \ref{spectrum} shows the time evolution of the other occupied
proton states 
in our example of a $^{16}\textrm {O}$+$^{16}{\textrm
O}$ collision.  As for the $1s_{1/2}$ levels above, we again see the
splitting of the asymptotically degenerate states $1p_{3/2}$ and $1p_{1/2}$ 
with increasing
interaction.  The difference between adiabatic and boost-invariant
energies indicates the degree of internal excitation. It can be very
different for the different states. It is, e.~g., somewhat surprising
that the lower state (left panels) acquires excitation
rather early while some higher states wait for much longer.
The lower panels of figure \ref{spectrum} show the occupation
probabilities (\ref{eq:adiabocc}). They start at unity as they should
for a yet unexcited state where adiabatic and boost-invariant states
are still identical. Their subsequent decrease reflects the degree of
intrinsic excitation. The two indicators for intrinsic excitation,
occupation probability and difference
$\varepsilon^\mathrm{(inv)}-\varepsilon^\mathrm{(adia)}$, agree nicely
for all states shown.
The adiabatic energies in the upper panels of figure \ref{spectrum} show the
interesting phenomenon of level crossings, quite similar to that observed in former studies
using deformed shell models
\cite{Park80,Park82,Cas85a,Maruhn72}. The arrows connect these points
with the lower panels where the occupation probabilities seem to jump. But that
is merely a labeling effect when sorting the states always according to
adiabatic energy. Following the states diabatically through the crossings would
produce smooth evolution of energies and occupation probabilities. Such a
diabatic tracking, however, is only possible if we ignore pairing, as we do
here. Inclusion of pairing would smoothen the crossings and enforce adiabatic
tracking with subsequent ``smooth jumps'' in the observables.

\subsection{Landau-Zener effect}

\begin{figure*}
\epsfxsize=13cm 
\centerline{\epsffile{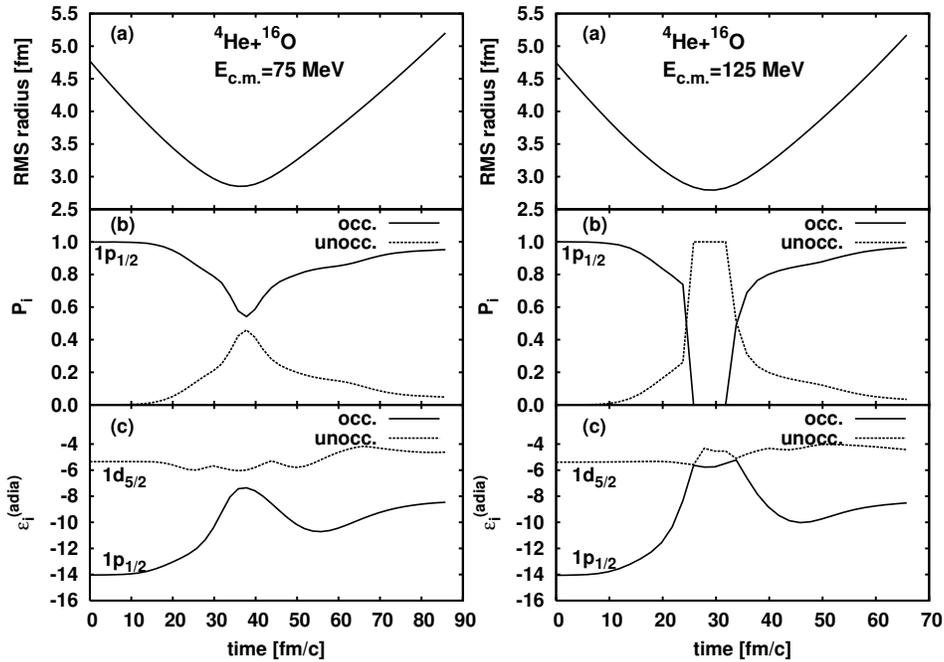}}
\caption{\label{LZ} 
Time evolution of the central collision of
$^{4}\textrm{He}$+$^{16}{\textrm O}$ with c.m. energies of 75
MeV and 125 MeV. (a) rms radius for the
colliding system as a function of time; (b) adiabatic occupation
probability $P_i$ of the last occupied neutron state; and (c) 
interacting single-particle adiabatic states. 
The labels of the single-particle states at t=0 are 
near the left axis.}
\end{figure*}

As mentioned in the introduction, reactions of complex
many-body systems like nuclei or molecules often produce
the Landau-Zener effect.
It happens at level crossings with only small coupling between
the two energetically close levels. There is a competition
between the speed with which the levels evolve and the
time necessary for the rearrangement of the occupations.
Very slow motion leaves sufficient time such that
always the energetically lower level is fully occupied.
That is the adiabatic limit. Increased velocity causes
transitions (= diabaticity) where the occupation partially
crosses over into the then higher level, thus turning collective
energy to internal excitation. That is the much celebrated
Landau-Zener effect.
In this section, we will employ the boost-invariant and adiabatic
states as analyzing tools to a study of the Landau-Zener effect in
a self-consistent mean-field description of
heavy-ion collisions. For this purpose an asymmetric reaction is more
appropriate and we select a head-on collision of $^{4}\textrm
{He}$+$^{16}{\textrm O}$.

Fig.~\ref{LZ}(a) shows the root-mean-square (RMS) radius of the colliding system as a function of
time.  It is clear that both the collisions with 75 and 125 MeV are deep-inelastic scattering.  The
adiabatic occupation probability as defined in Eq.(\ref{eq:adiabocc}) for the last occupied neutron
state labelled at initial stage as $1p_{1/2}$ is presented in Fig.~\ref{LZ}(b). The assignment
``occ.'' and ``unocc.'' in Fig.~\ref{LZ} refers to the situation in the initial stage, while in the
colliding region the adiabatic occupation probability gives information on the actual occupation.

The occupation probability is, of course, nearly unity during the
initial stage of nearly free translation and starts decreasing as the
two colliding nuclei approach each other. It then increases again and
finally returns to nearly unity, not quite attaining it owing to the 
small transfer and evaporation probabilities
after the separation. The corresponding
adiabatic single-particle energies of the highest-lying occupied and
the lowest-lying unoccupied neutron states, 
labelled at the initial stage as $1p_{1/2}$ and $1d_{5/2}$ respectively, are shown in
Fig.~\ref{LZ}(c). In the initial stage of dynamic propagation, the
adiabatic single-particle energies are almost the same as those of the
static ground states (realized exactly at $t=0$).

This behavior is easy to understand since the boost-invariant and
adiabatic states only reflect the excitation and interaction of
colliding nuclei. The two adiabatic states display the feature of
avoided crossing around the smallest distance of two colliding
nuclei. The same feature also appears in the adiabatic occupation
probability. The right panel with larger bombarding energy of 125~MeV
shows that the mixing of occupied and unoccupied components in the
colliding stage becomes much stronger such that the occupied and
unoccupied states are exchanged completely. Since the two interacting
single-particle states belong to the nucleus $^{16}$O, we find that
two neutrons are excited from the uppermost occupied state to the
lowest unoccupied state.  This excitation is activated gradually with
increasing incident energy. This is a clear signal of a nuclear
Landau-Zener transition in the TDHF description of a deep inelastic
collision.

\section{\label{level5} Summary}

In this work, we have constructed a boost-invariant single-particle
Hamiltonian to eliminate the dynamically induced variances coming from
the local velocity field in TDHF. For the case of free translational
motion of a nucleus, the boost-invariant Hamiltonian produces
eigenstates which have zero dynamical variances and reproduce the
stationary single-particle energies. In the case of a reaction, true
intrinsic excitations take place and the TDHF states do not remain
eigenstates of the boost-invariant Hamiltonian anymore. Their
variances then become a measure of intrinsic excitation and the
expectation values still remain useful measures of single particle
energies. Moreover, the eigenstates of the boost-invariant Hamiltonian
can be considered as the (instantaneous) adiabatic states which
contain no flow. The relation between boost-invariant and adiabatic
single-particle energies is also related to the intrinsic excitation
energy, similar to the energy variances. As a further measure of
excitation, we introduce occupation probabilities, i.~e., the probabilities
to find a given adiabatic state within the space of occupied TDHF
states. Adiabatic states and occupation probabilities serve as
analyzing tools, e.~g., to investigate the nuclear Landau-Zener effect
within self-consistent mean-field models.
The scheme has been implemented numerically in fully three-dimensional
TDHF without any symmetry restrictions and with all the terms of the
Skyrme energy functional included. 

Two test cases of head-on collisions were considered, fusion of the
$^{16}$O+$^{16}$O system at low scattering energy and deep inelastic
scattering of $^{4}$He+$^{16}$O. The newly defined boost-invariant and
adiabatic single-particle energies show the expected behaviors.  For
the symmetric $^{16}$O+$^{16}$O system, the splitting of the
asymptotically degenerate levels in the interaction regime is clearly
seen.  In both cases, one finds the mutually complementing signals for
intrinsic excitation and, in particular, several nicely developed
level crossings with, in the case of $^{4}$He+$^{16}$O, all signatures
of a nuclear Landau-Zener effect. The trend from more
adiabatic evolution at low energies to clean diabatic transitions at
high collisional energy, e.~g., is clearly apparent.

These first results are very encouraging. The boost-invariant
Hamiltonian with its single-particle energies and the corresponding
adiabatic basis are promising tools for analyzing TDHF simulations of
heavy-ion reactions and understanding their relation to the other
widely used time-dependent method based on single-particle orbitals,
the expansion in the adiabatic basis.

\begin{acknowledgments}
Lu Guo acknowledges the support from the Alexander von Humboldt
Foundation.  We gratefully acknowledge support from the Frankfurt Center
for Scientific Computing. The work was also supported in part by the BMBF
under Contracts No. 06 F 131 and 06 ER 808.
\end{acknowledgments}

\bibliography{adia_tdhf}

\end{document}